# A Feature Selection Method that Controls the False Discovery Rate

Mehdi Rostami, Olli Saarela

November 7, 2023

**Abstract**

The problem of selecting a handful of truly relevant variables in supervised machine learning algorithms is a challenging problem in terms of untestable assumptions that must hold and unavailability of theoretical assurances that selection errors are under control. We propose a distribution-free feature selection method, referred to as Data Splitting Selection (DSS) which controls the False Discovery Rate (FDR) of feature selection while obtaining a high power. Another version of DSS is proposed with a higher power which "almost" controls FDR. No assumptions are made on the distribution of the response or on the joint distribution of the features. Extensive simulation is performed to compare the performance of the proposed methods with the existing ones.

## 1 Introduction

The problem of feature selection has received a huge amount of attention due to the explosion of high dimensional data sets. In the supervised machine learning algorithms (Friedman et al., 2001), it is often important to choose a handful of features, among possibly hundreds or thousands, which are truly associated with the outcome of interest and are used in the model construction (Candes et al., 2018). Reasons include reducing noise to the model results due to an excessive number of irrelevant features in the model, and making the model more understandable (Cai et al., 2018). However, selecting a handful of true predictors in supervised machine learning algorithms is a challenging problem in terms of verifiable assumptions that must hold and unavailability of theoretical assurances that selection errors are under control. As an example, one of the most commonly used feature selection methods in classical ML is the LASSO (Tibshirani, 1996) which automatically shrinks regression parameters toward zero and theoretically, should generate an exact zero for noise features. However, there is no guarantee that the selected features (features with non-zero effects) will be re-selected in independent studies and if either type I and/or II selection errors are controlled. By type I error we mean the probability of selecting noise features, and by type II error we mean the probability of not selecting true or relevant features.

We consider the problem of carrying out a feature selection algorithm that control type I and II errors of selection for low to high dimensional settings without making strong assumptions that are hard to verify or satisfy in practice.

There is a wealth of literature on selection methods that control some form of type I/II errors with their own advantages and shortcomings. For example, the Benjamini-Hochberg (BH) (Benjamini and Hochberg, 1995) powerful method of p-value adjustment [1] that controls the False Discovery Rate (FDR) is one of the most practiced methods in the regression models and multiple hypothesis testings, especially in Genomics (Korthauer et al., 2019). FDR is defined as the average ratio of false discoveries over all discoveries, $\mathbb{E}[\frac{V}{\max(1,R)}]$, with $V$ and $R$ being the number of false discoveries and total number of discoveries, respectively. Since the regression coefficients determine the strength of association between the features and the outcome. The BH method is only applicable to methods where pvalues are defined and p-values are exact and it assumes exchangeability of hypotheses (Efron, 2012). That is, the modifications of BH that relax this assumption are less powerful (Efron, 2012).

The multiple comparison problem is more general than hypotheses testing and is raised in feature selection literature, too. A recent and powerful feature selection method called the knockoff method, developed by Barber et al. (2015) theoretically controls FDR in finite samples. The original method, referred to as fixed-X knockoff assumed normality, linearity and homoscedasticity assumptions (Barber et al., 2015), and the other version model-X knockoff relaxes these assumptions, yet it relies on the joint probability distribution of features to be known (Candes et al., 2018). The knockoff method is motivated by the permutation method (Molnar, 2020) where one permutes the features and compares the feature importance before and after the permutation so that large difference evidences against the null hypothesis. A drawback of the permutation method is that it destroys the correlation structure between the inputs by permuting the individual features. Barber and his colleagues aimed, in their paper (Barber et al., 2015), to rectify this by constructing a multivariate copy of the features (here referred to as knockoff features) with the condition that the correlation structure remains unchanged, but each knockoff feature is independent of the original feature. Barber et. al. have applied knockoff in high dimensional settings (Barber and Candès, 2016) and have discovered a lower bound on the power of the knockoff procedure (Fan et al., 2019) using data splitting ideas. An earlier and similar work to knockoff was introduced by (Wu et al., 2007) which creates pseudo-variables instead of knockoff features.

These aforementioned methods are among the methods that select features without deploying any resampling methods. To address this issue, Yoav Benjamini and Ruth Heller (Benjamini and Heller, 2008) introduced a method called the Partial Conjunction Hypothesis (PCH) that applies the BH procedure on different studies (or splits of the original data) and selects hypotheses that are rejected more than a pre-specified number of studies. Another method called the Stability Selection (SS) method selects features that have high probability of being associated with the outcome at least for one hyper-parameter (Meinshausen and Bühlmann, 2010). This method provably controls average number of false discoveries, which is different from FDR.

These methods make certain assumptions about the relationship between features, or distributional assumptions on the outcome of interest. The X-version of knockoff method is computationally expensive and not easily scalable to high dimensions. We propose the 'Data Splitting Selection' (DSS) method and its modification 'Multiple Sampling Selection' (MSS). These methods regard a feature as important to be selected if it shows a strong association with the outcome in multiple data sets. In DSS, the data set is randomly split into two parts and the association between features and response are investigated in both parts. In a supervised problem, DSS selects a feature if it shows a strong association with the outcome in both parts of the data. MSS is similar in nature but uses sub-sampling with replacement instead of data splitting. As it will be shown, no assumptions on the models, estimation processes or joint distribution of features are required. In addition, DSS and MSS (especially when parallelized) are computationally cheap and are competitive in terms of selection power in high dimensional data. DSS and MSS also insure the replicability of the findings by utilizing two or more samples rather than one.

---

[1] This method is applied as a feature selection procedure in GLMs. In fact, the null hypothesis $H_0 : \beta_j = 0$ equivalent to $H_0 : j^{th}$ feature is null.

In this draft we interchangeably use the words "selection" and "discovery." Needless to mention that assuming the null hypotheses to be $j^{th}$ feature is not important (or is null), for all $j$, "selection" is equivalent to rejection of a hypothesis whose feature is null. Thus, the selected features are said to be non-null. Similarly, by "the selection method has high power" we mean the method can select truly important features.

In this paper, we will not make any assumption about the relationship between the inputs and the outcome, or about the joint distribution of the inputs. However, our selection algorithms need the data model or the machine learning algorithm to have a measure of importance for the features in advance. The over-the-shelf method of measuring feature importance is to measure the absolute difference between the loss function in the absence and presence of a feature. Regression parameters in regression models, lasso and ridge methods and the "feature importance" in tree-based methods such as random forests, and virtually any other feature importance measure whose large magnitude is evidence against the null hypothesis (1) can be used in our feature selection algorithm.

The main contribution of this article is the introduction of two model-agnostic feature selection methods referred to as Data Splitting Selection (DSS) and Multiple Sampling Selection (MSS) with the property of controlling FDR while having a large power. The methods can also be applied to unsupervised algorithms, when the relationship between features is of interest, such as a graphical model.

To be more rigorous, in a supervised problem, with outcome $y$ and features $X_1, ..., X_p$, we define the $j^{th}$ null hypothesis as

$$H_0 : \text{Variable } X_j \text{ is not associated with variable y.} \qquad (1)$$

In a regression model, where $\beta_j$ is the coefficient of $X_j$, this hypothesis can be stated as

$$H_0 : \beta_j = 0.$$

Type I error is the probability of rejecting a null hypothesis while it is actually true. Type II error is the probability of not rejecting a null hypothesis while it is actually false and should be rejected. The power of a selection method is one minus its type II error, the probability of correctly rejecting a null hypothesis. Any feature selection method is prone to these errors due to randomness in the observed data. Controlling the False Discovery Rate (FDR) (Benjamini and Hochberg, 1995), the average of false discoveries among selected features, to be smaller than a pre-specified nominal FDR is one way to control the type I error. Controling the probability of at least one false discovery, referred to as the Family-Wise Error Rate (FWER), is another way of controlling for type II error (Efron, 2012). FDR-type selection (or hypothesis testing) methods are usually more powerful than FWER-type.

The organization of the rest of the article is as follows: The DSS and MSS methods are introduced in Section 2. Section 3 is devoted to the simulations and the comparison of the DSS, MSS, knockoff and Benjamini-Hochberg (BH) methods. In section 4, we demonstrate the applicability of the methods Which methods? by applying them to an undirected graphical model to decide which nodes should be connected. Theoretical results are provided in the Appendix.

**Algorithm 1** The DSS algorithm for feature selection.

1. Split the Data set $D$ randomly into mutually exclusive $D_1$ and $D_2$
2. Standardized the input features in both sets
3. Trained models on both $D_1$ and $D_2$
4. Calculate new feature importance statistics (FIs) taken from both splits (2)
5. Calculate the threshold from FIs.
6. Select the features whose FIs are larger than the threshold.

## 2 Data Splitting Selection

The Algorithm 1 describes the steps of the Data Splitting Selection (DSS) algorithm. The procedure start with randomly splitting the data to two independent sets, the training and validation sets: $D = D^{tr} \cup D^v$ and standardize the inputs to have zero mean and unit variance in each part separately. To evaluate the association between the features and the outcome, a model is fitted to both data sets such as penalized or ordinary regression model or a tree-based machine learning algorithm equipped with feature importance statistics (Friedman et al., 2001). Let $w_j$ be the population parameter which shows true correlation/association between the $j^{\text{th}}$ feature and the outcome and $\hat{w}_j^{tr}$ and $\hat{w}_j^v$ be statistics estimating this parameter on the training and validation sets, respectively.

We define the test statistics $Z_j^{tr} = |\hat{w}_j^{tr}|$, and $Z_j^v = |\hat{w}_j^v|, \forall j = 1, ..., p$ (which also show the importance of each feature in both splits). Then the $j^{\text{th}}$ (new) feature importance ($FI_j$) is defined as

$$FI_j = Z_j^{tr} I_{\pm 1}(Z_j^v > \tau), \qquad (2)$$

where $I_{\pm 1}(\mathcal{A}) = I(\mathcal{A}) - I(\mathcal{A}^c)$ with $I(.)$ being the indicator function; $Z_j^{tr}$ and $Z_j^v$ are the statistics corresponding to $j^{\text{th}}$ input evaluated on the training and validation sets, respectively. $\tau$ is a positive scalar which may be fixed or estimated from the validation data.

A threshold based on the calculated feature importance statistics $FI$'s is calculated:

$$T = \min\left\{t \in \left\{Z_1^{tr}, Z_2^{tr}, ..., Z_p^{tr}\right\} : \frac{\#\{j : FI_j \leq -t\}}{\max(\#\{j : FI_j \geq t\}, 1)} \leq q\right\} =$$
$$\min\left\{t \in \left\{Z_1^{tr}, Z_2^{tr}, ..., Z_p^{tr}\right\} : \frac{\#\{j : Z_j^{tr} \geq t, \ Z_j^v < \tau\}}{\max(\#\{j : Z_j^{tr} \geq t, \ Z_j^v \geq \tau\}, 1)} \leq q\right\} \quad (3)$$

where $\tau$ is a quantity independent of $t$. Also, define $T = \infty$ if the set is empty. $T$ defines a threshold on the largeness of feature importance statistics. This threshold is similar to and, in fact, is motivated by the one defined in (Barber and Candès, 2016).

The final step of the selection procedure is to select features with high association with the response in both splits, that is features whose feature importance $FI$ are larger than the threshold: $S = \{j : FI_j \geq T\} = \{j : Z_j^{tr} \geq T, \ Z_j^v \geq \tau\}$. The parameter $\tau$ can be determined using an elbow-type method or a greedy submodular optimization.

Note that DSS will have zero power if $T = \infty$. This can happen when $\nexists \ t \in \left\{Z_1^{tr}, Z_2^{tr}, ..., Z_p^{tr}\right\}$ such that $\frac{\#\{j:FI_j \leq -t\}}{\max(\#\{j:FI_j \geq t\}, 1)} \leq q$. This can be visualized in right panels of figure 2: When ratio of statistics in region R1 (purple) over those in region R2 (green) is not small compared to $q$.

## 2.1 FDR Control

Considering the definition of the tests statistics on training and validation sets, it is clear that $Z_j^{tr} \perp Z_j^v$, that is $Z_j^{tr}$ and $Z_j^v$ are independent. This property helps prove the main result of this manuscript:

**Theorem 1.** *Let the subjects be independent and* $\mathbf{w} = (w_1, w_2, ..., w_p)$ *be vector of associations between features and the response. Also, the statistics should be so that larger values are evidence against null hypotheses. If $\tau$ is chosen so that*

$$Pr_{H_{0j}}(Z_j^v \geq \tau) \leq Pr_{H_{0j}}(Z_j^v < \tau), \tag{4}$$

*with $Pr_{H_{0j}}(.)$ being the probability of the event under null ($w_j = 0$), the False Discovery Rate (FDR) is controlled up to the nominal level $q$, that is*

$$\text{FDR} = \mathbb{E}\left[\frac{\#\{j : w_j = 0 \text{ and } j \in S\}}{\max(\#\{j : j \in S\}, 1)}\right] \leq q. \tag{5}$$

The proofs are provided in the appendix. Some plain text here to explain the theorem, and its significance.

**Estimate $\tau$**

The value of the hyperparameter $\tau$ balances the trade-off between type I and II errors; $\tau \approx 0$ will give 100% power and a high type I error, and as $\tau \to +\infty$, the power and type I error approach zero.

In the above theorem, the hyperparameter $\tau$ is assumed to be a fixed positive quantity. In an ideal world that we know the exact distribution of test statistics under the null, the hyperparameter $\tau$ can be found with the best trade-off. For example, in a linear regression where the distribution of $\hat{w}_j$ for null features is $N \sim \mathcal{N}(0, \sigma_0^2)$, we can select $\tau = x_{75\%}$, where $x_{75\%}$ is the $75^{th}$ percentile of the null distribution $N$. Figure 1 visualizes the white area under the normal curve.

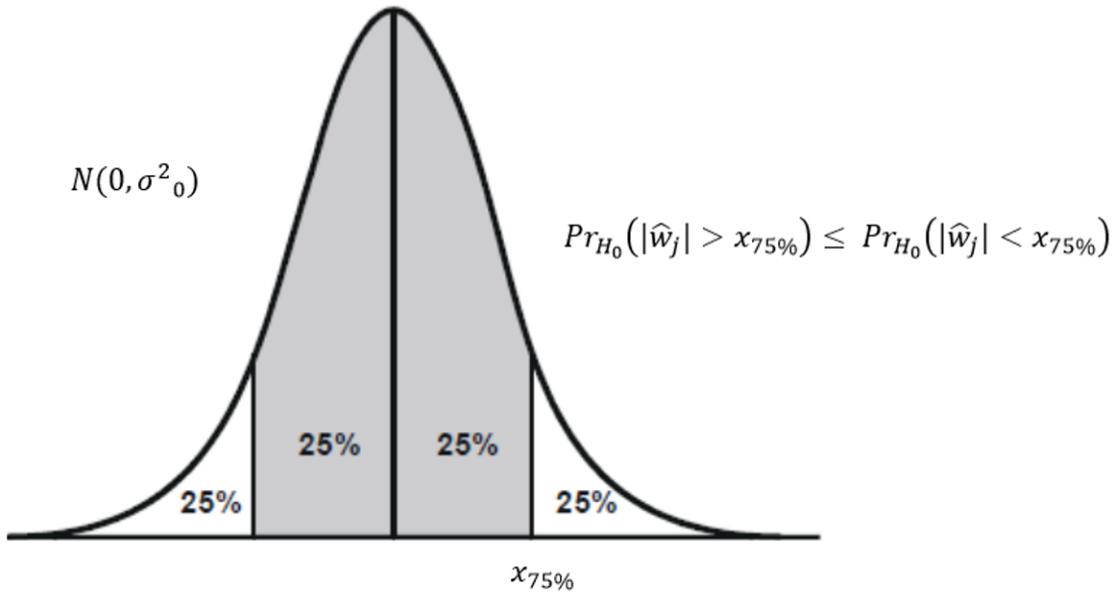

Figure 1: Best choice of $\tau$ in a particular case when the true distribution of the parameter estimators are normally distributed.

In practice, the exact value of the hyperparameter $\tau$ is hard or impossible to know and it must be estimated using the data at hand. However, any method of estimation will result in some level of type I and II errors. We introduce a method, here referred to as the elbow method. To estimate $\tau$, we calculate and plot the empirical distribution $F$ of $\{Z_j\}_{j=1}^p$. The estimate is some $Z_k$ ($k \in \{1, 2, ..., p\}$) so that $F(_k)$ is furthest from the line that connects minimum and maximum of $Z_j$'s. The two figures on the left side of Figure 2 illustrate the elbow method in a simulation setting, where the test statistics are absolute value of the slopes in an ordinary (top figures) and $L_1$ penalized (bottom figures) regression models. The right two panels show the regions of True and False discoveries. The green region is the selection region (where $Z_j^{tr} \geq T$ and $Z_j^v \geq \tau$). Red +'s indicate Type II errors, green −'s demonstrate the true null acceptance (not rejecting null hypotheses), green +'s show the power, and green −'s refer to the false rejections; the latter is not visible in the pictures.

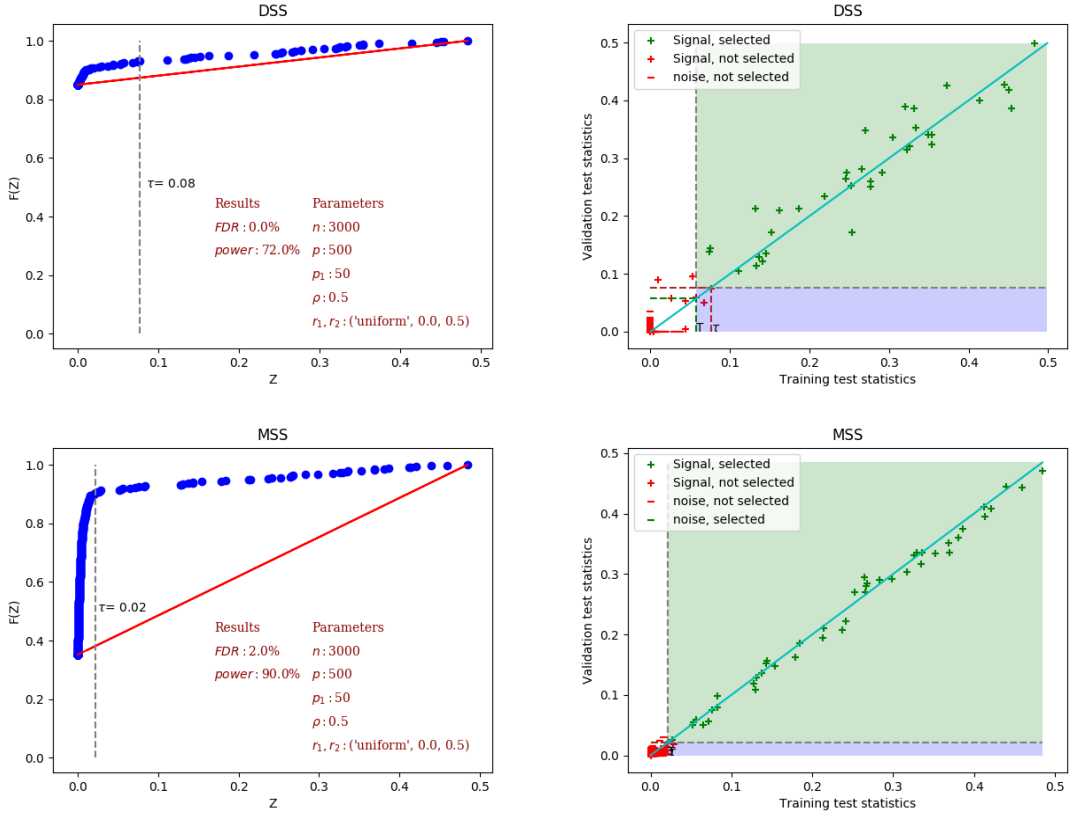

Figure 2: An illustration of the elbow method (the figures on the left) and type I and II error trade-off (the figures on the right). The top figures are associated with the ordinary regression model and those at the bottom to the $L_1$ penalized regression model. The green region in the right panels show the selection region (where $Z_j^{tr} \geq T$ and $Z_j^v \geq \tau$). Green + shows True positives (selected signals), red + shows sitives (unselected signals), green − shows True Negative (unselected noises), and red − show the False Negative (selected noises).

## 2.2 Multiple Sampling Selection

The DSS method would have a high power with minimal type I error if the truth about the distribution of $Z_j$'s were known. However, the elbow method does not lead to a high power. This is due to the reduced sample sizes in DSS, plus the fact that the influential data points can fall in either of the two sets and result in the rejection of a true signal. This shortcoming of DSS can be remedied by sub-sampling rather than splitting. This method is referred to as Multiple Sampling Selection (MSS) (Algorithm 2) and the process starts with randomly selecting $k + k'$ sub-samples with replacement from the original data $D$, that is $D_1^*, D_2^*, ..., D_{k+k'}^*$. Let $\hat{w}_{1,j}^\star, \hat{w}_{2,j}^\star, ..., \hat{w}_{k+k',j}^\star$ be the estimates of the association between the $j^{\text{th}}$ variable and outcome in the $k + k'$ data sets. Define

$$\begin{aligned} Z_j^{tr} &:= \frac{1}{k}\big|\hat{w}_{1,j}^\star + \hat{w}_{2,j}^\star + ... + \hat{w}_{k,j}^\star\big| \\ Z_j^{v} &:= \frac{1}{k'}\big|\hat{w}_{k+1,j}^\star + \hat{w}_{k+2,j}^\star + ... + \hat{w}_{k+k',j}^\star\big|, \end{aligned} \quad (6)$$

for $j \in \{1, 2, ..., p\}$ and where $|.|$ is the absolute value function.

**Corollary 1.** *Let $Z_j^{tr}$ and $Z_j^{v}$ be statistics defined as* (6). *Adopting the selection procedure of DSS and letting $\tau > 0$ be such that*

$$Pr_{H_{0j}}(Z_j^v \geq \tau | Z_j^{tr} \geq T) \leq Pr_{H_{0j}}(Z_j^v < \tau | Z_j^{tr} \geq T), \quad (7)$$

*FDR is controlled up to the nominal level $q$, that is*

$$\text{FDR} = \mathbb{E}\left[\frac{\#\{j : w_j = 0 \text{ and } j \in S\}}{\max(\#\{j : j \in S\}, 1)}\right] \leq q.$$

---

**Algorithm 2** The MSS algorithm for feature selection.

1. Randomly select $k + k'$ sub-samples $(D_1^*, D_2^*, ..., D_{k+k'}^*)$ with replacement from $D$.
2. Standardize the input features in each $D_j^*$
3. Train $k + k'$ models on all $D_j^*$s
4. Calculate new feature importance statistics (FIs) using all estimates (6)
5. Calculate the threshold
6. Select the features whose FIs are larger than the threshold.

---

## 3 Simulations

To investigate and understand DSS and MSS, we have compared them with Benjamini-Hochberg (BH)(Benjamini and Hochberg, 1995), knockoff (Barber and Candes, 2016), and the Stability Selection (SS) (Meinshausen and Bühlmann, 2010) methods in a simulations study. The simulation scenarios vary based on sparsity levels, range of (true) signal sizes and correlation size among the inputs as they impact the performance of the selection method. Initial simulations indicated that the performance of the algorithms is massively impacted by the signal sizes. Thus a further investigation is made where the signal sizes are too small. Thus the simulations are divided into two classes of Small-to-large signal size scenarios, and Small signal size scenarios.

### 3.1 Small-to-Large Signal Sizes

The factors that usually impact the performance of feature selection methods are the correlations among inputs, signal sizes, and sparsity (or the dimension of the true predictive features). To compare the performance of the proposed methods with the popular existing methods (Knockoff, BH, SS), we have simulated 100 different simulation scenarios with 3000 data points, and 1000 inputs based on different values of these factors:

- Numbers for true signals ($p_1$) are 20, 50, 100;
- Correlation parameter value ($\rho$) between inputs are 0, 0.5, 0.7, 0.9;
- Range of magnitude of true signals ($r_1, r_2$): (0, 0.5), (0.5, 1), (1, 2).

Other parameters of the simulation scenarios are fixed as

- $\mathbf{X} \sim \mathcal{N}(\mathbf{0}_p, \Sigma)$, with $\Sigma_{kj} = \rho^{j-k}$, which means in each scenario, only a handful of the features are correlated;
- $\mathbf{y} = \mathbf{Xw} + \boldsymbol{\epsilon}$  where $\boldsymbol{\epsilon} \sim \mathcal{N}(\mathbf{0}_n, \mathbf{I}_n)$, and $\mathbf{w} \in (r_1, r_2)$;
- The tests statistics are defined to be absolute value of regression coefficients with lasso penalty;
- Feature importance for knockoff method is defined $FI_{knock} = \max(|\hat{w}|, |\hat{\tilde{w}}|) * \text{sign}(|\hat{w}| - |\hat{\tilde{w}}|)$, where $\hat{w}$ and $\hat{\tilde{w}}$ are regression parameter estimates corresponding to the inputs and knockoff features, respectively, and $FI_{knock}$ stands for the feature importance defined in the knockoff method.
- Feature importance for the DSS method is defined $FI = Z^{tr} I_{\pm}(Z^v > \tau)$, where $Z^{tr} = |\hat{w}^{tr}|$ evaluated on half of the data and $Z^v = |\hat{w}^v|$ evaluated on the other half. And similarly, for the MSS.
- The threshold $\tau$ was estimated according to the elbow of the empirical distribution of the regression parameter estimates on the training data as shown in figure 2.
- In the Stability Selection (SS) method, we have chosen the threshold to be 0.7 for the probability that a feature should be selected.

## 3.2  Simulation Results

Figures 3 and 4 illustrate the difference between FDR and power of the five methods DSS, MSS, knockoff, BH and SS methods [2]. In all of the scenarios, the FDR of the DSS is very close to zero, and at the same time, if the signals are not too small, its power is as large as the other two methods. The figures also show that the strength of the correlation between the features does not significantly influence the performance of DSS and BH, but MSS, Knockoff and SS methods produce more false discoveries with higher correlation, and for extreme correlation levels, SS generates false discoveries more than twice of the nominal level. In terms of power, all the methods result in very high power when signals are strong, but when true signals are between 0 and .5, the power of all the methods are substantially smaller, and shrink even further when cross-correlation among features goes up. The worst performance is for DSS, especially when correlations are very small. Except for the DSS, the power is not influenced by the sparsity level so that the methods are consistent for the same signal size across varying sparsity levels. DSS outperforms MSS and the two other methods in terms of FDR, but that is misleading in cases of small signal sizes since the power of DSS is a lot smaller than that of MSS.

---
[2] The power is the sample average of proportion of true discoveries and FDR is the sample average of FDPs.

In terms of run time, the BH is the fastest, and DSS, MSS, SS and knockoff are in the next places, respectively. Due to the construction of knockoff feature which involves optimization techniques, plus feeding all the original and knockoff features in the model for parameter estimation, this method runs slower than the other four.

### 3.2.1 Skewness of the Distribution of False Discovery Proportion

A selection method which always picks most of the true signals is a better method than one that picks most of the true signals in some data sets, and no true signal in others. The FDR which is the expected value of the False Discovery Proportion (FDP) can be a misleading measure giving a high average in both of the above cases (Efron et al., 2007). To investigate the performance of our proposed selection methods, we have looked at the distribution of FDP (Figures 5 and 6). It can be seen that, when signal sizes are small and sparsity is high, DSS has the lowest FDR, but the power goes down, too. The best performance is for MSS which, generally, induces higher power and acceptably low FDR compared to the other four methods in almost all of the cases.

### 3.2.2 Small Effect Sizes

In this set of scenarios, we let the true signals to be very small, and vary the other scenario parameters as before:

- True signals $\in (0, .2)$
- $\rho \in \{0.0, 0.5, 0.7, 0.9\}$
- $p_1 \in \{20, 50, 100\}$

Figures in 7 illustrate the average and summary statistics of false and true discovery proportions (FDP and power). The DSS fails to detect the signals most of the time, especially when the number of true signals is small (where $\tau$ is the elbow of the empirical distribution function.) The FDR of MSS goes slightly above the nominal level in extreme correlation levels, and FDR of SS increases about twice or more than twice of the nominal level. The knockoff and MSS have relatively lower FDPs (and FDR), but the power of knockoff can be very low, particularly when sparsity is high. Examining the box plots 8 shows that the MSS method is the most stable and consistent method among others.

## 4 Real Data Experiment

## 5 Discussion

In this article, we have proposed a new selection method which controls the false discoveries through controlling FDR, with no major assumption that confines its applicability. We have performed a thorough simulation study of the performance of our proposed methods in terms of type I and II errors. The theoretical results also prove that FDR is controlled, and if the right hyperparameter is chosen, the power of the algorithm will be high, too. As the economist Thomas Sowell has argued: "There are no solutions, there are only trade-offs; and you try to get the best trade-off you can get, that's all you can hope for." In this discussion, the hyperparameter $\tau$ is the trade-off between the type I and II errors, and we have shown that the elbow method of choosing $\tau$ in MSS does result in high power and controlled FDR.

The DSS method is closely related to the replicability analysis introduced in (Benjamini and Heller, 2008; Benjamini et al., 2009) where the authors define null hypotheses according to the replicability of findings in multiple samples. These methods rely on independence or positive dependence of hypotheses. Our works is also similar to Stability Selection (Meinshausen and Bühlmann, 2010). Stability Selection (SS) is a re-sampling method aiming to estimate the probability that a feature is important for a grid of hyperparameters. The SS method is different from our methods in multiple ways. For example, in the lasso problem, we use cross-validation to find a hyper-parameter, and use the estimates of regression parameters to define feature importance, while in SS, we choose a grid of hyperparameters in advance and run simulations for each value of hyperparameter to estimate the proportion of the number of times that the feature is selected. SS also depends on a hyperparameter that is not clear how to fix, though we set it 0.7 in our simulations as recommended by the inventors of SS. Importantly, DSS/MSS controls the FDR while SS does not.

Although it is observed that DSS has low FDR, in many scenarios it does not have high power. The competing methods might select more true features, but their analyses are based on distributional assumptions which are impossible or hard to verify, or if there are verifying methods, these methods may not have enough power to detect the null hypothesis that the assumptions are satisfied (Freedman, 2009). If the model assumptions are true, this method is not optimal and other parametric methods may be more appropriate. Otherwise, generally speaking, a non-parametric method may result in less susceptible trustworthy results. Furthermore, to the best of our knowledge, there is no general method controlling for type I error which can be applicable to any machine learning method such as Gradient Boosting Machines.

The naive random sampling for DSS and MSS is inappropriate when data are clustered or correlated such as in autocorrelated spatial and time series data. One method is to randomly select clusters or bins (for time series) that are independent and consider all of the subjects in the clusters or bins. The two-stage sampling is also recommended when data matrix is sparse. In addition, it is possible that the data contains outliers or extreme values. This may cause one of the splits containing the extreme values and affect the estimates and the inference. This issue can partially be addressed in MSS when we lump multiple estimates from different random samples.

In summary, we have proposed an assumption-free method which is fast and more accurate in making type I error at the expense of losing some minor power. These methods can be added to the toolbox of data analysts when the distributional assumptions are doubtful for feature selections.

# 6    Appendix A

**Lemma 1.** *Let*

$$Pr_{H_{0j}}(FI_j \geq T) \leq Pr_{H_{0j}}(FI_j \leq -T) \tag{8}$$

*hold for feature importance statistics, where $Pr_{H_{0j}}(.)$ is the probability of the event under null $(w_j = 0)$. Hence,*

$$\#\{j: \ w_j = 0, \ FI_j \geq T\} \stackrel{\mathrm{d}}{\leq} \#\{j: \ w_j = 0, \ FI_j \leq -T\}. \tag{9}$$

*Proof.* First, according to (Torres, 2013), for the random variables $X_1, Y_1, X_2, Y_2$, we have

$$X_j \stackrel{\mathrm{d}}{\leq} Y_j \quad \text{iff} \quad \exists \ X'_j \stackrel{\mathrm{d}}{=} X_j, \ Y'_j \stackrel{\mathrm{d}}{=} Y_j, \ \text{and} \ M_j > 0, \ s.t. \ X'_j + M_j \stackrel{\mathrm{d}}{=} Y'_j, \tag{10}$$

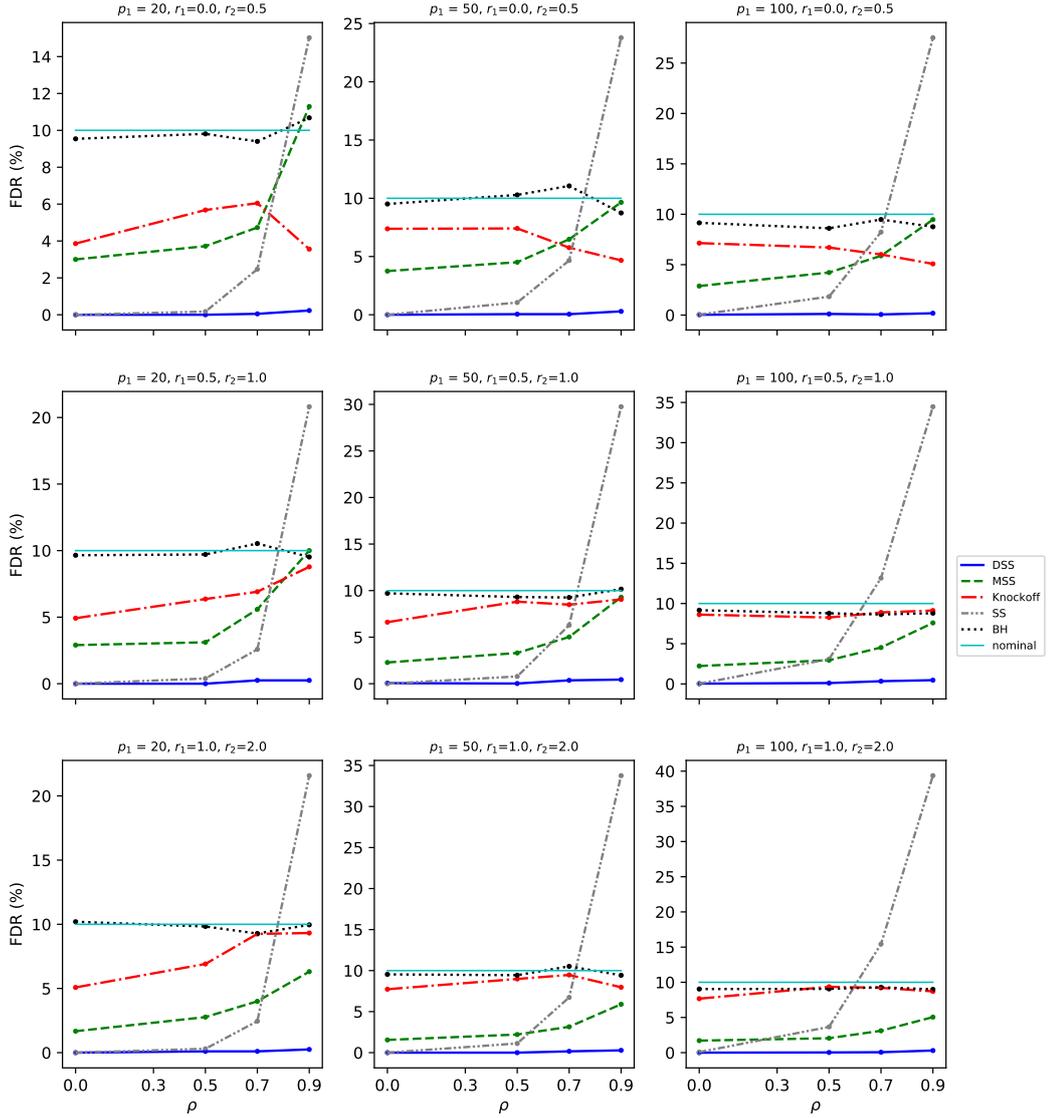

Figure 3: FDR (%) of the 4 methods, DSS, MSS, knockoff, SS and BH. The nominal level is 10% ($q = .1$). All the methods control FDR up to 10% except SS for very high correlation among inputs. Among all, DSS and MSS have overall lowest FDR and BH the highest, while SS has very low FDR for small or correlations but it performs poorly for high correlation cases. However, DSS and BH are completely consistent in false discovery rate control across different spectrum of correlation and signal size.

for $j = 1, 2$. Therefore, if $X_1 \overset{d}{\leq} Y_1$ and $X_2 \overset{d}{\leq} Y_2$, then

$$Y_1 + Y_2 \overset{d}{=} Y_1' + Y_2' \overset{d}{=} X_1' + M_1 + X_2' + M_2 \overset{d}{=} X_1 + M_1 + X_2 + M_2 \overset{d}{\geq} X_1 + X_2. \qquad (11)$$

This result can inductively generalize to more than 2 random variables.

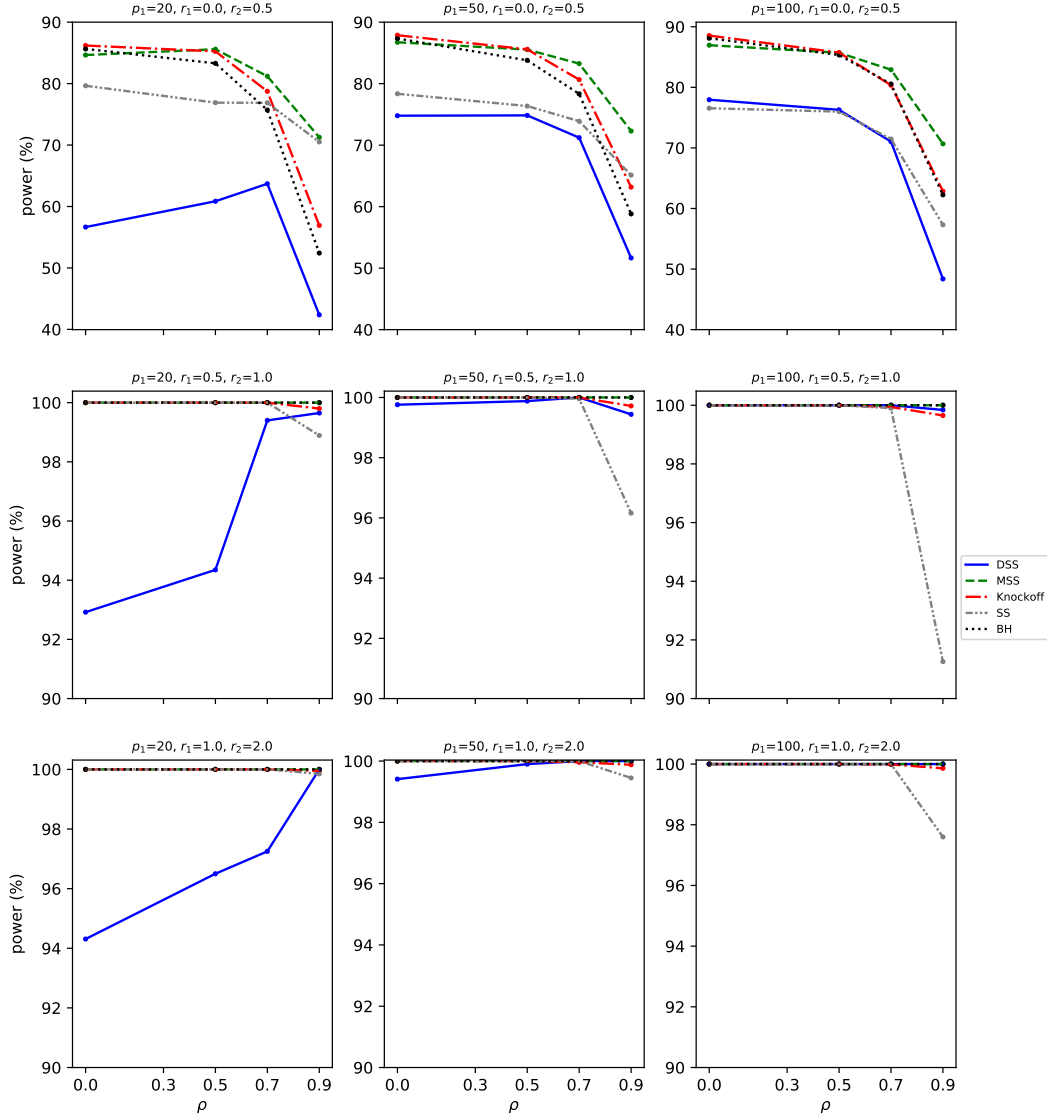

Figure 4: The power (%) of the 4 methods, DSS, MSS, knockoff, SS and BH. When signal size is high, all the methods result in very high power (above 90%). However, for small true effect sizes, higher correlation results in lower power. Among all, roughly speaking, MSS has higher power than other methods.

Now, let $X_j \sim \text{Bernoulli}(p_j)$ and $Y_j \sim \text{Bernoulli}(q_j)$, with $p_j = Pr_{H_{0j}}(FI_j \geq T)$ and $q_j = Pr_{H_{0j}}(FI_j \leq -T)$. By the assumption, we have that $p_j \leq q_j$ which implies $X_j \stackrel{d}{\leq} Y_j$, $\forall$ null $j$. Hence, the inequality

$$\#\{j: w_j = 0, FI_j \geq T\} = \sum_j X_j \stackrel{d}{\leq} \sum_j Y_j = \#\{j: w_j = 0, FI_j \leq -T\}$$

holds. □

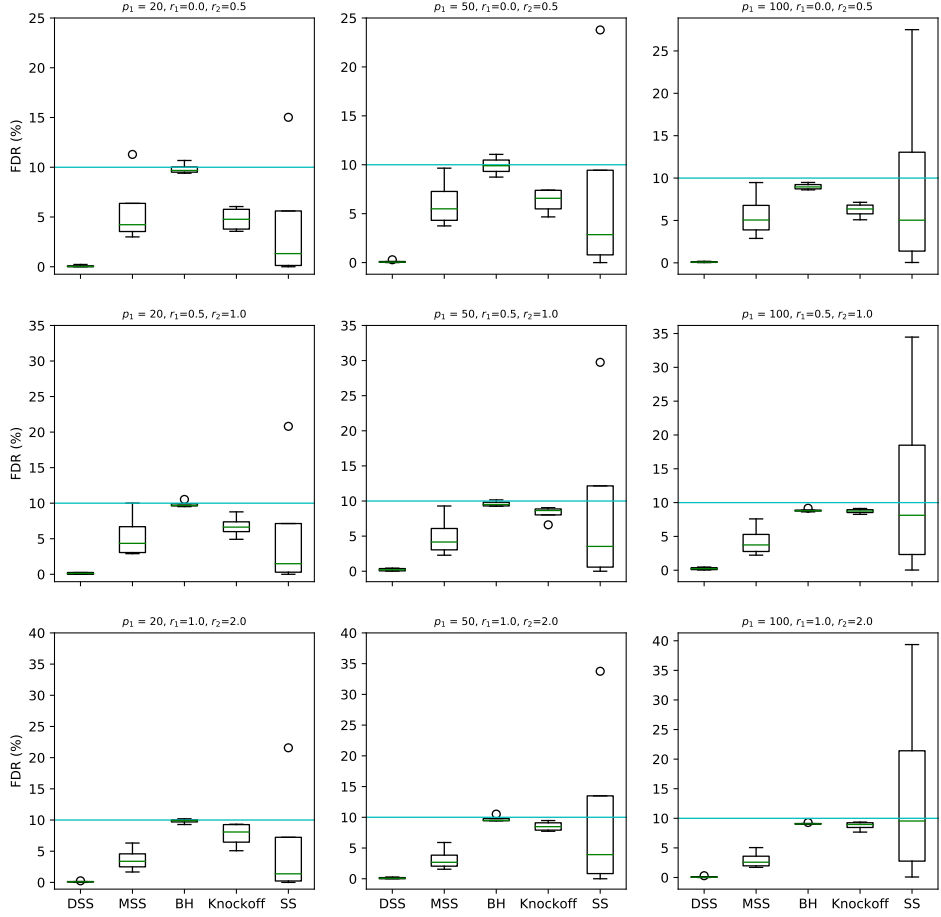

Figure 5: The distribution of FDP (%) induced by DSS, MSS, BH, knockoff and SS methods in various settings. DSS and MSS have lower FDPs (which is in fact misleading for DSS by looking at the power distribution.)

**Theorem 1.** *For T selected as*

$$T = \min\left\{t \in \left\{Z_1^{tr}, Z_2^{tr}, ..., Z_p^{tr}\right\} : \frac{\#\{j : FI_j \leq -t\}}{\max(\#\{j : FI_j \geq t\}, 1)} \leq q\right\} \tag{12}$$

*and $\tau$ that satisfies*

$$Pr_{H_{0j}}(Z_j^v \geq \tau) < Pr_{H_{0j}}(Z_j^v < \tau), \tag{13}$$

*the FDR of the selection is controlled up to nominal error $q$:*

$$\text{FDR} = \mathbb{E}\left[\frac{\#\{j : w_j = 0 \text{ and } j \in S\}}{\max(\#\{j : j \in S\}, 1)}\right] \leq q.$$

*Proof.* First, we show the assumption of theorem 1 is satisfied. Assuming 13, for an arbitrary $t > 0$ and a fixed $\tau > 0$, we have

$$Pr_{H_{0j}}(FI_j \geq t) \stackrel{\text{def}}{=} Pr_{H_{0j}}(Z_j^{tr} I_{\pm}(Z_j^v \geq \tau) \geq t) = Pr_{H_{0j}}(Z_j^{tr} \geq t \ \& \ Z_j^v \geq \tau) \stackrel{\text{indep}}{=}$$
$$Pr_{H_{0j}}(Z_j^{tr} \geq t) Pr_{H_{0j}}(Z_j^v \geq \tau) \stackrel{\text{ineq.13}}{\leq} Pr_{H_{0j}}(Z_j^{tr} \geq t) Pr_{H_{0j}}(Z_j^v < \tau) \stackrel{\text{indep}}{=}$$
$$Pr_{H_{0j}}(Z_j^{tr} \geq t \ \& \ Z_j^v < \tau) \stackrel{\text{def}}{=} Pr_{H_{0j}}(Z_j^{tr} I_{\pm}(Z_j^v \geq \tau) < -t) \stackrel{\text{def}}{=} Pr_{H_{0j}}(FI_j < -t), \quad (14)$$

where indep stands for independence of the training and validation sets. Now recall that

$$T = \min \left\{ t \in \{Z_1^{tr}, ..., Z_p^{tr}\} : \ \frac{\#\{j : FI_j < -t\}}{\max(\#\{j : FI_j \geq t\}, 1)} \leq q \right\}. \quad (15)$$

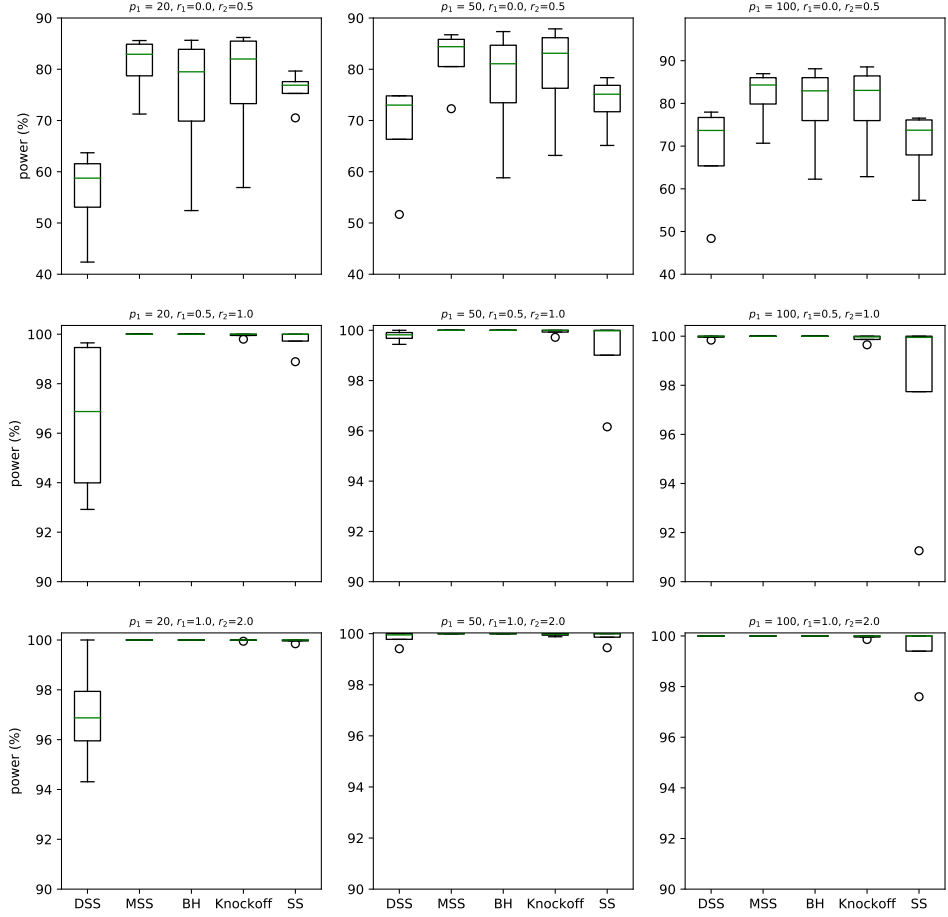

Figure 6: The distribution of power (%) induced by DSS, MSS, BH, knockoff, and SS methods in various settings. Except DSS and SS, the other methods induce high power in almost all of data sets generated especially when signal sizes are high. In small signal sizes, MSS has higher power overall.

If $T = \infty$, no feature is selected and FDR is zero. Assuming a finite $T$ exists we replace $t = T$ in 14, and thus $Pr_{H_{0j}}(FI_j \geq T) \leq Pr_{H_{0j}}(FI_j \leq -T)$.

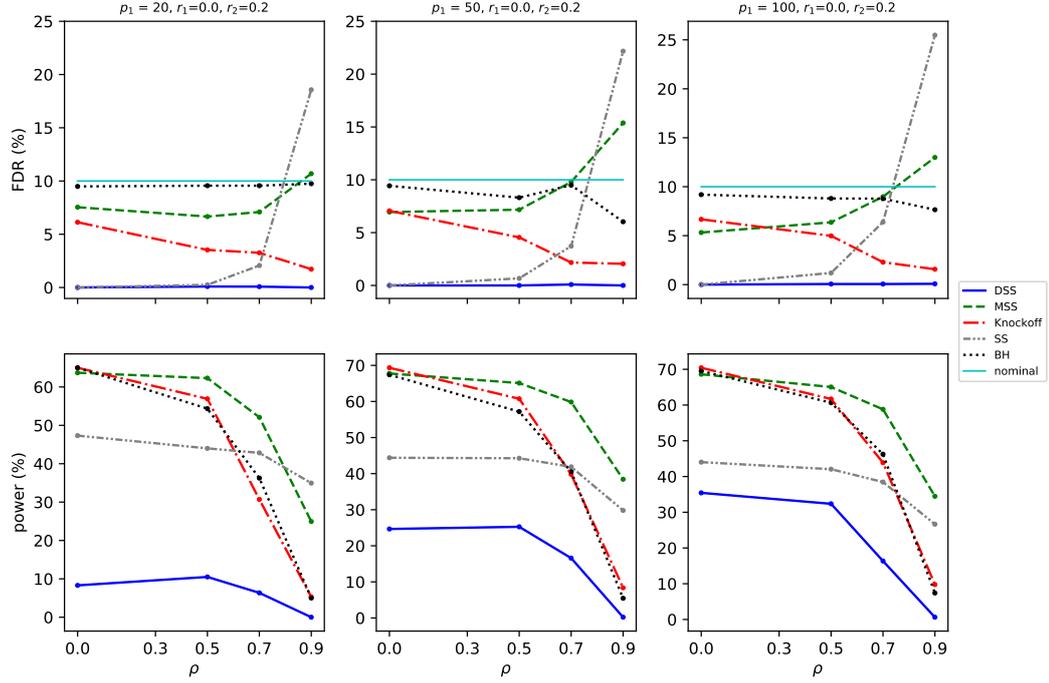

Figure 7: The FDR and power (%) of the methods when signals are small. Comparing FDR and power of all the methods in the three scenarios, it can be observed that MSS is superior with having relatively low FDR at the same time of having high power than other methods.

Therefore, applying the result of lemma 1 yields:

$$\text{FDP} = \frac{\#\{j:\ w_j = 0,\ FI_j \geq T\}}{max(1, \#\{j:\ FI_j \geq T\})} \stackrel{d}{\leq} \frac{\#\{j:\ w_j = 0,\ FI_j < -T\}}{max(1, \#\{j:\ FI_j \geq T\})} \leq$$

$$\frac{\#\{j:\ FI_j < -T\}}{max(1, \#\{j:\ FI_j \geq T\})} \stackrel{3}{\leq} q,$$

which implies

$$\text{FDR} = \mathbb{E}\,\text{FDP} \leq q.$$

□

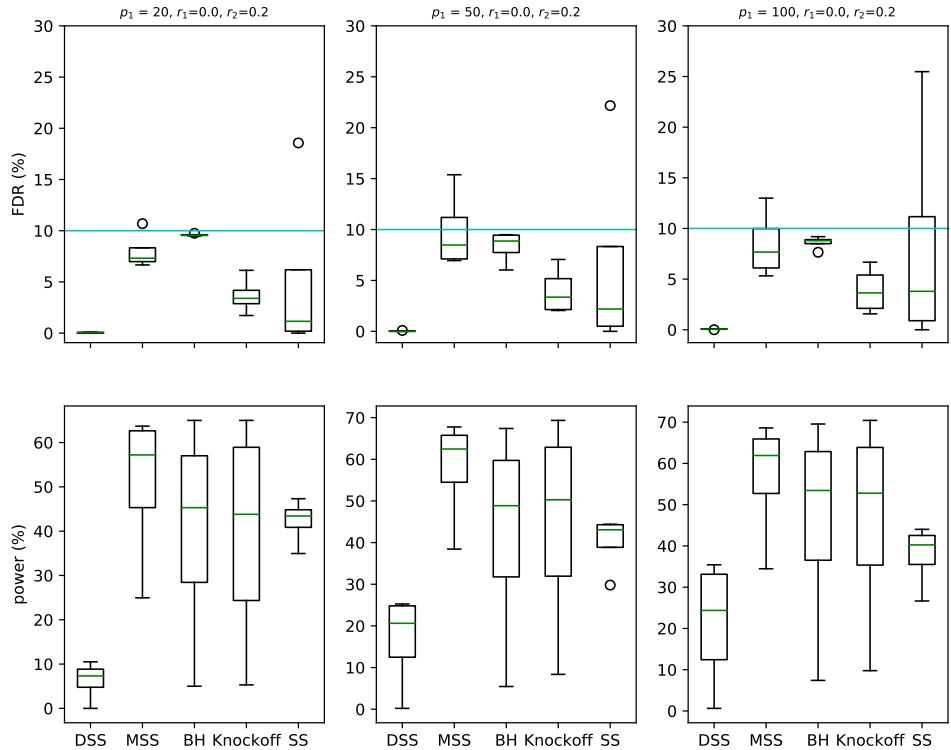

Figure 8: The FDP and power (%) of the methods when signals are small. Comparing distributions of FDP and power of all methods in the three scenarios, it can be observed that MSS is superior with having low FDP at the same time of having high power than other methods. Note that in a case that knockoff has lower FDR than MSS, it has lower power, too.